# *In Silico* Investigations on the Potential Inhibitors for COVID-19 Protease[*]

Ambrish Kumar Srivastava[1], Abhishek Kumar[2, 4], Gargi Tiwari[3], Ratnesh Kumar[2], Neeraj Misra[2]

[1]*Department of Physics, Deen Dayal Upadhyaya Gorakhpur University, Gorakhpur (U.P.) India*

[2]*Department of Physics, University of Lucknow, Lucknow (U.P.) India*

[3]*Department of Physics, Patna University, Patna (Bihar) India*

[4]Corresponding author: abhishekphy91@gmail.com (A. Kumar)

## Abstract

A novel strain of coronavirus, namely, COVID-19 has been identified in Wuhan city of China in December 2019. There are no specific therapies available and investigations regarding the treatment of the COVID-19 are still lacking. This prompted us to perform a preliminary *in silico* study on the COVID-19 protease with anti-malarial compounds in the search of potential inhibitor. We have calculated log $P$ and log $S$ values in addition to molecular docking and PASS predictions. Among the seven studied compounds, mepacrine appears as the potential inhibitor of the COVID-19 followed by chloroquine, hydroxychloroquine and phomarin. Therefore, these anti-malarial drugs may be potential drug candidate for the treatment of this novel coronavirus. A detailed analysis on these inhibitors is currently in progress and clinical studies are invited to investigate their potential medicinal use for the COVID-19.

**Keywords:** COVID-19; Inhibitors; Anti-malarial drugs; Mepacrine; Chloroquine; Molecular Docking.

**Introduction**

At the beginning of this year, the coronavirus disease (COVID-19) was recognized in Wuhan, China [1]. Subsequently, the COVID-19 started spreading across the globe, putting the whole world on high alert [2–5]. This led to 655 total cases and 18 deaths all over the world including China and 9 other countries till 23 January 2020 [6]. In India, the first case of the COVID-19 was reported in Kerala on 30 January 2020. As of 20 March 2020, there are 230 cases and 4 deaths as reported by the Ministry of Health and Family Welfare, Government of India [7]. According to a report of the world health organization (WHO) dated 17 March 2020, there are a total of 0.18 million cases of the COVID-19 worldwide, causing almost 7.5 thousand deaths and counting [8].

Coronaviruses infect humans and vertebrate animals, affecting their respiratory, digestive, liver and central nervous systems [9]. Since the inception of the COVID-19 at the end of 2019, the continuous efforts have been made in the research and development of the diagnostics, therapeutics and vaccines for this novel coronavirus [10]. Based on the results of some clinical trials, it has been reported [11] that chloroquine phosphate, an anti-malarial drug, has a certain curative effect on the COVID-19. In particular, chloroquine phosphate is recommended to treat COVID-19 associated pneumonia in larger populations in the future. Subsequently, another research based on clinical trials suggested [12] that hydroxychloroquine added with azithromycin is very effective in the treatment of the COVID-19. This motivated us to perform a systematic study on some anti-malarial drugs using molecular docking and reinvestigate their biological activities and pharmacological effects. Such studies become important as they offer some insights into structure-based drug design. Recently, Lin et al. [13] have reported the structure-based stabilization of non-native protein-protein interactions of the coronavirus in antiviral drug design. However, there exists no systematic study on the inhibition of the coronavirus by anti-malarial drugs to the best of our knowledge. Therefore, we believe that this study should offer better insights into the binding and interaction of anti-malarial drugs with the COVID-19 receptor.

**Methodology**

To identify the potential binding sites for anti-malarial compounds, we have been performed an automated *in silico* molecular-docking procedure using the SwissDock web server [14, 15], which is based on the docking algorithm EADock ESS. We have used the main protease in corona viruses as a potential target protein for COVID-19 obtained from the RCSB protein data bank (PDB ID: 6LU7) [16]. The processed coordinates file for each of the ligands and COVID-19 protein (6LU7) has been uploaded, and docking was performed using the 'Accurate' parameter option, which is most exhaustive in terms of the binding modes sampled.

We have also calculated lipophilicity (log $P$) and aqueous solubility (log $S$) ALOGPS 2.1 program [17]. This program is based on electrotopological state indices and associative neural network modeling developed by Tetko et al. [18]. The log $P$ and log $S$ are two very important parameters for quantitative structure-property relationship (QSPR) studies. In order to explore and predict the pharmacological effects and biological activities of molecules, we have used PASS software [19]. The PASS predicts 900 pharmacological effects, molecular mechanisms of action, mutagenicity, carcinogenicity, teratogenicity and embryotoxicity.

**Results and Discussion**

We have chosen seven popular anti-malarial compounds as a ligand to the COVID-19 protease namely mepacrine (**1**), chloroquine (**2**), quinine (**3**), hydroxychloroquine (**4**), artemisinin (**5**), phomarin (**6**) and proguanil (**7**). The molecular structures of these drugs are displayed in Fig.

1. In order to assess and compare their biological activity, we have computed their log *P* and log *S* values as listed in Table 1. Log *P* is closely associated with the transport property of drugs and their interaction with receptors whereas log *S* is an important factor affecting its bioavailability. One can see that the log *P* values of compounds 1-7 lie in the range 6.13-1.90. Chloroquine (**2**) has a log *P* value of 5.28 followed by hydroxychloroquine (**4**) with a log *P* of 3.87. However, mepacrine (**1**) has the highest log *P* of 6.13 among all seven compounds studied in this work. Likewise, the log *S* value of these anti-malarial compounds ranges between -2.35 and -5.22. In general, about 85% of drugs have log *S* values in the range of -1 to -5, a few have values below -5 but virtually none have values below -6 [20]. All log *S* lies between -1 and -5, except mepacrine (**1**) having log *S* slightly below -5. These log *P* and log *S* values suggest that the drug molecules may easily diffuse across the cell membranes, as their organic (lipid) solubility is quite large.

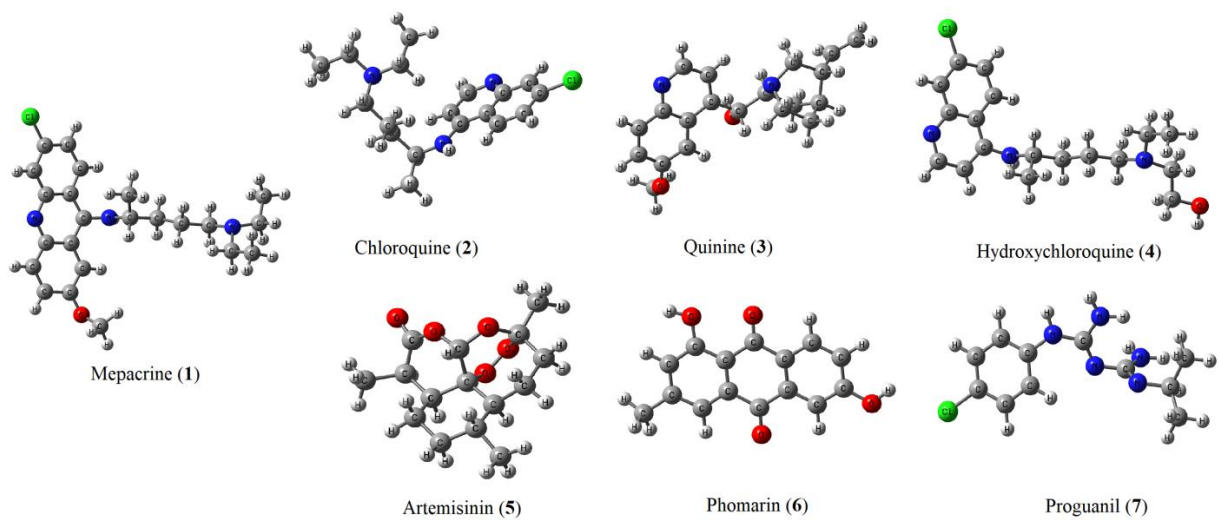

Fig. 1. Molecular structures of anti-malarial compounds as possible inhibitors for 6LU7 (COVID-19 protease).

This study has been focused on the main protease in the novel corona virus as a potential target protein for COVID-19 treatment. The 6LU7 [21] is the main protease found in the COVID-19, which has been deposited as a PDB file in early February 2020. Therefore, we have chosen 6LU7 as a potential target for molecular docking studies. The molecular docking calculations have been performed as blind, i.e., covered the entire protein surface, not any specific region of the protein as the binding pocket in order to avoid sampling bias. The output clusters have been obtained after each docking step and ranked according to the full fitness (FF) score by the SwissDock algorithm. A greater negative FF score suggests a more favorable binding mode between ligand and receptor with a better fit.

Table 1: Calculated parameters of anti malarial compounds as possible COVID-19 inhibitors.

| S.No. | Anti malarial compounds | Log $P$ | Log $S$ | Residue & Bond lengths | Binding affinity (kcal/mol) | FF score |
|---|---|---|---|---|---|---|
| 1 | Mepacrine ($C_{23}H_{30}ClN_3O$) | 6.13 | -5.22 | GLU166 (2.663 Å) | -8.89 | -1189.0 |
| 2 | Chloroquine ($C_{18}H_{26}ClN_3$) | 5.28 | -4.26 | GLY143 (2.321 Å) | -8.15 | -1208.0 |
| 3 | Quinine ($C_{20}H_{24}N_2O_2$) | 2.82 | -2.99 | HSD163 (2.377 Å) | -7.77 | -1144.0 |
| 4 | Hydroxychloroquine ($C_{18}H_{26}ClN_3O$) | 3.87 | -4.11 | PHE140 (2.501 Å) | -7.62 | -1184.0 |
| 5 | Artemisinin ($C_{15}H_{22}O_5$) | 2.52 | -2.35 | GLY143 (2.447 Å, 2.369 Å) | -7.34 | -1187.0 |
| 6 | Phomarin ($C_{15}H_{10}O_4$) | 3.04 | -3.35 | GLY143 (2.462 Å) GLU166 (2.607 Å) | -7.13 | -1192.0 |
| 7 | Proguanil ($C_{11}H_{16}ClN_5$) | 1.90 | -2.95 | LEU141 (2.262 Å) | -6.69 | -1347.0 |

The results of molecular docking are displayed in Fig. 2. The docking parameters such as binding energy, FF score along with the amino acids (residues) found in the active site pockets of 6LU7 have been also listed in Table 1. The binding energies (affinities) obtained from docking of 6LU7 with ligands mepacrine (**1**), chloroquine (**2**), quinine (**3**), hydroxychloroquine (**4**), artemisinin (**5**), phomarin (**6**) and proguanil (**7**) are -8.86, -8.15, -7.77, -7.62, -7.34, -7.13 and -6.69 kcal/mol respectively. The affinity of drug compounds depends on the type of bonding that occurs with the active site of the protein. The results of docking analyses show the mepacrine (**1**) forms H-bond (bond length = 2.663 Å) with the glutamate (GLU-166), a polar amino acid. On the contrary, chloroquine (**2**) forms H-bond (bond length = 2.321 Å) with the glycine (GLY-143), a non-polar amino acid. Unlike **2**, quinine (**3**) and hydroxychloroquine (**4**) form H-bonds with the histidine (HSD-163), a polar amino acid (bond length = 2.377 Å) and phenylalanine (PHE-140), an aromatic amino acid (bond length = 2.501 Å). Artemisnin ((**5**) forms two H-bonds of bond lengths 2.447 Å and 2.369 Å with the same amino acid (GLY-143). Phomarin (**6**) also forms two H-bonds, one with the GLY-143 (bond length = 2.462 Å) and another with the GLU-166 (bond length = 2.607 Å). Proguanil (**7**) forms H-bond of 2.262 Å with the leucine (LEU-141), a non-polar amino acid. The binding modes of different anti-malarial compounds with the 6LU7 receptor have been shown in Fig. 2 as well. Thus, our docking analyses suggest that the COVID-19 protease (6LU7) can be inhibited by anti-malarial drug compounds. Based on the binding affinity, the inhibition potential of these compounds can be ranked as; **1** > **2** > **3** > **4** > **5** > **6** > **7**. Combining the results of log *P* and log *S* with the docking, we can expect that the compounds **1**, **2**, **4** and **6** should behave as potential inhibitors of the COVID-19.

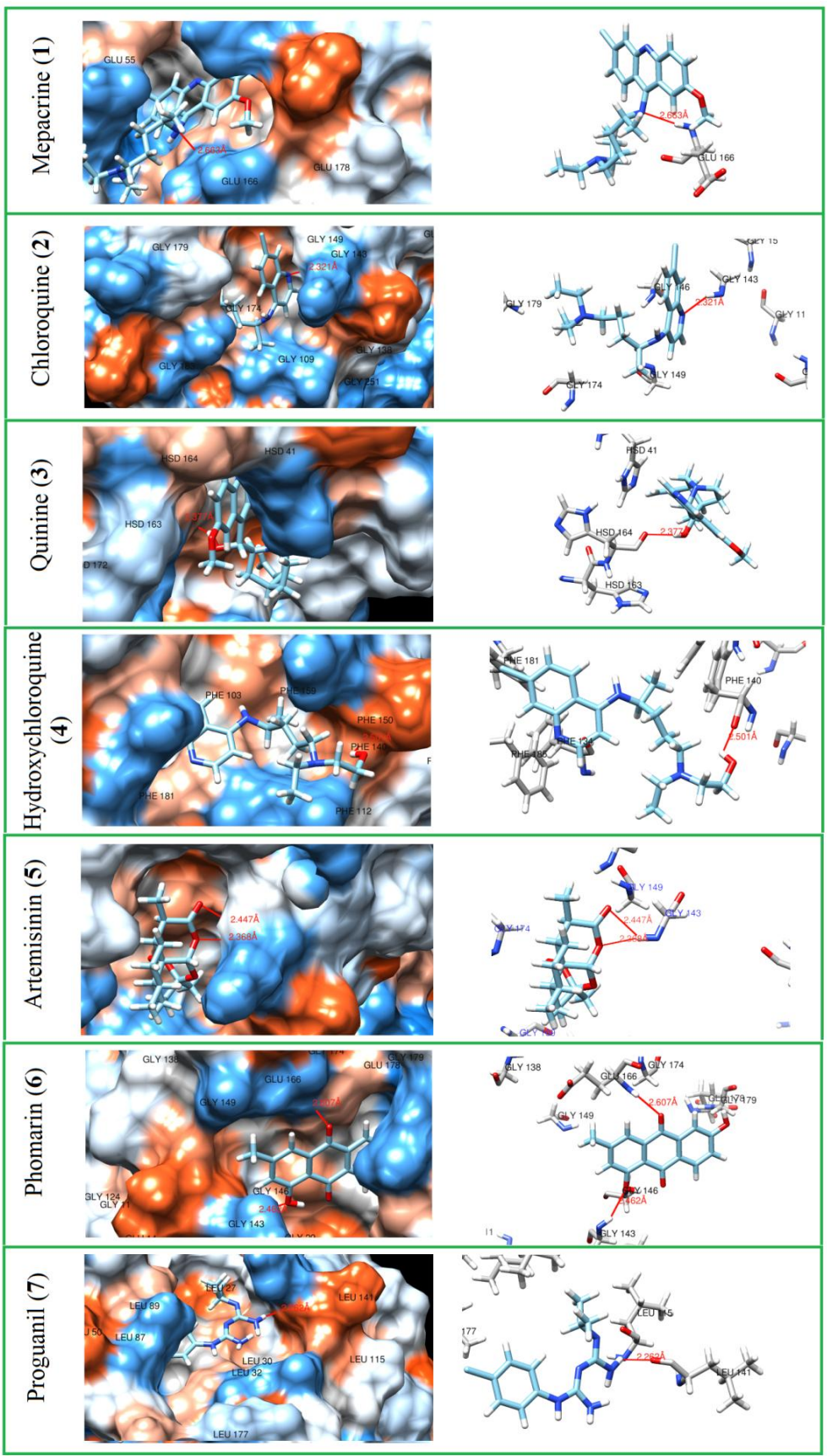

Fig. 2. Active binding sites of 6LU7 (COVID-19 protease) explored by molecular docking.

For the purpose of drug design, it is desirable to analyze the detailed biological activities or possible pharmacological effects along with toxicity or possible adverse effects of the drug compounds. Therefore, we have predicted activity spectra of anti malarial compounds using the PASS software. This prediction is based on the study of structure activity relationships (SAR) for the training set including about 50,000 drugs, drug-candidates and lead compounds whose biological activity has been reported experimentally. An average accuracy of prediction in leave-one-out cross-validation is about 85% [19]. In supplementary Table S1, we have listed the pharmacological effects as well as adverse affects of these compounds with Pa > 70%. This is to ensure that the molecules will most likely exhibit these activities in the experiment. Furthermore, the probability that the molecules have several closely analogous drugs is quite high. These results of PASS prediction might be useful during clinical trials and drug development stage of the COVID-19.

**Concluding Remarks**

We have performed a systematic study on anti-malarial compounds in the search of potential inhibitors for novel coronavirus, COVID-19 protease. Based on the binding affinity, mepacrine appears as the most powerful inhibitor among seven compounds studied here. However, its log $P$ is quite high and log $S$ value is very low. Other potential inhibitors of COVID-19 protease include chloroquine, hydroxychloroquine and phomarin. Therefore, a detailed analysis on these drugs is required. Meanwhile, we suggest some clinic trials of these compounds or their suitable combinations. Our PASS predictions may be useful for clinical trials of these inhibitors of COVID-19. Further studies in these directions are in progress in our lab and shall be reported shortly.

**Acknowledgement**

AKS acknowledges University Grants Commission (UGC), New Delhi, India for startup grant [30-466/2019(BSR)]. Authors are also thankful to Prof. S. N. Tiwari for useful discussions.